\documentclass[preprint,12pt]{elsarticle}

\usepackage{amssymb}
\usepackage{graphicx}
\usepackage{multicol,multirow}
\usepackage{amsmath,amssymb,amsfonts}
\usepackage{mathrsfs}
\usepackage{amsthm}
\usepackage{makecell}
\usepackage{rotating}
\usepackage{appendix}
\usepackage[numbers]{natbib}
\usepackage{ifpdf}
\usepackage[T1]{fontenc}
\usepackage{newtxtext}
\usepackage{newtxmath}
\usepackage{textcomp}
\usepackage{xcolor}
\usepackage{lscape}
\usepackage{lipsum}
\usepackage{bbm}
\usepackage{array}
\usepackage{booktabs} 
\usepackage{caption} 
\usepackage{xcolor} 

\usepackage{mathtools}
\usepackage{stmaryrd}

\usepackage{booktabs}
\usepackage{pgfplots}
\pgfplotsset{compat=1.14}
\usepackage{anyfontsize}
\usepackage{pifont}

\usepackage{eqparbox}
\usepackage{array}
\usetikzlibrary{backgrounds}
\usepackage{soul}
\usepackage{icomma}
\usepackage{marvosym}
\usepackage{textcomp}
\usepackage{color, colortbl}
\usetikzlibrary{arrows}

\usepackage{bbm}
\usepackage{geometry}
\usepackage[colorlinks,allcolors=blue]{hyperref}
\usepackage{caption}
\usepackage{subcaption}

\theoremstyle{definition}

\numberwithin{equation}{section}

\begin{document}

\begin{frontmatter}



\title{Age Group Sensitivity Analysis in Age Stratified Epidemic Models: \\Investigating the Impact of Contact Matrix Structure}


\author[bolyai,natlab]{Zsolt Vizi}

\author[bolyai]{Evans Kiptoo Korir}

\author[bolyai]{Norbert Bogya}

\author[bolyai]{Csaba Rosztóczy}

\author[bolyai]{Géza Makay}

\author[bolyai,wigner]{Péter Boldog}

\affiliation[bolyai]{organization={Bolyai Institute, University of Szeged},
            addressline={Aradi vértanúk tere 1}, 
            postcode={6720}, 
            state={Szeged},
            country={Hungary}}

\affiliation[natlab]{organization={National Laboratory for Health Security},
            addressline={Aradi vértanúk tere 1}, 
            postcode={6720}, 
            state={Szeged},
            country={Hungary}}

\affiliation[wigner]{organization={Wigner Research Centre for Physics},
            addressline={Konkoly-Thege Miklós út 29-33}, 
            postcode={1121}, 
            state={Budapest},
            country={Hungary}}

\begin{abstract}
Understanding the role of different age groups in disease transmission is crucial for designing effective intervention strategies. 
The contact matrix is a key parameter in age-structured epidemic models, which defines the interaction structure between age groups. However, accurately estimating contact matrices is challenging, as different age groups respond differently to surveys and are accessible through various channels. 
This variability introduces significant epistemic uncertainty in epidemic models.

In this study, we introduce the Age Group Sensitivity Analysis (AGSA) method, a novel framework for assessing the impact of age-structured contact patterns on epidemic outcomes. 
Our approach integrates age-stratified epidemic models with Latin Hypercube Sampling (LHS) and the Partial Rank Correlation Coefficient (PRCC) method, enabling a systematic sensitivity analysis of age-specific interactions. 
Additionally, we propose a new sensitivity aggregation technique that quantifies each age group's contribution to key epidemic parameters.

AGSA helps pinpoint the age groups to which the model is most sensitive by identifying those that introduce the greatest epistemic uncertainty. 
This allows for targeted data collection efforts, focusing surveys and empirical studies on the most influential age groups to improve model accuracy. 
As a result, AGSA can enhance epidemic forecasting and inform the design of more effective and efficient public health interventions.
\end{abstract}


\end{frontmatter}



\section{Introduction}

Mathematical models are extensively employed to understand and predict the spread of infectious diseases within populations. 
Compartmental models are a well-established and frequently applied method in mathematical epidemiology. In this approach, the population is divided into compartments (e.g., susceptible, infectious, recovered) that are significant in terms of the disease, and a system of coupled differential equations models the transitions between them. 

Such models effectively capture the dynamics of infectious diseases by simplifying complex interactions within a population. They allow researchers to quantify key epidemiological parameters, such as the basic reproduction number ($\mathcal{R}_0$) and the duration of outbreaks, providing insights into the progression and control of epidemics. Moreover, these models can be extended to include additional compartments or factors, such as vaccination, latency periods, or heterogeneous contact patterns, to reflect real-world scenarios better.

\subsection*{Uncertainty analysis and Latin Hypercube Sampling-Partial Rank Correlation Coefficient method}
In any mathematical model, the certainty of input factors such as parameters is often compromised by natural variations, measurement errors, or outdated measurement techniques \cite{alam2020parameter}.
Generally, parameters are derived from experimental data. In cases where parameter values are not estimated, they are set to plausible values or ranges based on statistical inference and expert judgment \cite{wu2013sensitivity}.
In epidemic modeling, a common and critical question is how altering the parameters affects the system’s temporal dynamics or descriptive quantities, such as the final epidemic size or the basic reproduction number. 

Sensitivity analysis, which performs Monte Carlo-style simulations to assess the impact of parameter changes, is a widely used approach to address this issue.
Sensitivity analysis measures how variations in input parameter values affect the model output \cite{wu2013sensitivity, blower1994sensitivity,  nsoesie2012sensitivity, marino2008methodology}.
This concept is used in various fields to identify (i) parameters that need further research to strengthen the knowledge base, (ii) insignificant parameters, (iii) inputs that majorly contribute to output variability, (iv) parameters highly correlated with the output \cite{hamby1994review}.
Sensitivity analysis approaches can be either local or global.

Local methods are straightforward and quick but only examine the effects of local parameter deviations or a selected trajectory in the parameter space.
Conversely, global techniques evaluate the entire parameter space along with interactions between parameters to identify all of the system’s critical points \cite{nsoesie2012sensitivity}.
Such techniques include variance-based methods (e.g., the Sobol' method and Fourier Amplitude Sensitivity Test) \cite{sobol1990sensitivity}, global screening methods (e.g., the Morris method) \cite{morris1991factorial}, and sampling-based methods (e.g., Monte Carlo filtering; scatter plots  \cite{friendly2005early}, Latin hypercube sampling with partial rank correlation coefficient index) \cite{helton2000sampling}.
This work focuses on the Latin Hypercube Sampling (LHS) combined with the Partial Rank Correlation Coefficient (PRCC).

LHS-PRCC sensitivity analysis is an advanced, efficient, and useful statistical technique commonly utilized in uncertainty analysis to explore a model's parameter space \cite{gomero2012latin}. The LHS/PRCC sensitivity analysis techniques have been applied to deterministic disease transmission models, as seen in \cite{alam2020parameter, wu2013sensitivity, blower1994sensitivity, nsoesie2012sensitivity, marino2008methodology}.

\subsection*{Age stratified models and social contact matrices}
Simple compartment models average out parameters characteristic of individuals, such as infectivity, recovery, or mortality rate, and consider the population as homogeneous. 
However, these parameters significantly depend on several factors, such as an individual's age, social behavior (contact pattern), or health status.
Moreover, different age groups interact in distinct ways within a population: the susceptibility to infection and the severity of disease may vary significantly by age.

Modeling the population in age-stratified compartments offers several advantages. 
By understanding how different age groups contribute to disease transmission and who is most at risk of severe outcomes, these models can guide decisions on which groups should be prioritized for vaccination to maximize public health benefits.
In scenarios where vaccines are limited, age-stratified models are crucial in determining optimal vaccination strategies or planning non-pharmaceutical interventions (NPIs). 
Policies like targeted lockdowns, age-specific social distancing measures, or selective school closures may be implemented based on model predictions, reducing the overall burden on society while maintaining effectiveness.
Thus, we may obtain a more realistic representation of how diseases spread across different population segments by stratifying the population into discrete age groups and incorporating \textit{age-dependent variations in parameters} and \textit{age-specific contact patterns} into the model.

Social contact patterns in different settings are categorized based on the age groups depicted by the contact matrix (CM). 
Correctly predicting infection dynamics and the effects of control strategies depends on precisely estimating the values of this contact matrix for the different age groups across various settings.
Several detailed studies have employed comprehensive quantitative approaches to evaluate social contact patterns in diverse settings and locations \cite{prem2021projecting, mossong2008social, fumanelli2012inferring, iozzi2010little, le2018characteristics, kiti2014quantifying, ajelli2017estimating, melegaro2017social, kumar2018interacts, read2014social, horby2011social, grijalva2015household, Koltai_reconst}. 
These studies have indicated consistent behavioral traits, i.e., assortative mixing; however, the frequency of contact, the extent of intergenerational mixing, and the nature of mixing can vary across age groups and attitudes \cite{korir2022clustering}.
Significant uncertainty hinders the precision of model output simulations over time for the transmission process, quantifying social contacts among different age groups in various settings.

This study aims to define a sensitivity measure for age-structured disease models in relation to the total number of contacts within each age group. 
This approach allows us to identify the age groups that have the most significant influence on the epidemic's key target parameters.
We use age-stratified epidemic models to achieve this and perform the LHS/PRCC technique and our novel sensitivity measure.

\bigskip

The subsequent chapters are organized as follows: the Methods section details the structure of contact matrices and the various elements of the proposed framework, including the main epidemic modeling concepts required for our work, Latin Hypercube Sampling (LHS), Partial Rank Correlation Coefficient (PRCC), the method for determining p-values, and the aggregation approach of PRCC values. 
In the Demonstrations section, we illustrate the proposed framework for different setups when parameterizing the epidemic model. 
Finally, we discuss the proposed framework's results and identify essential opportunities to modify and improve the presented methods.

\section{Methods}

\subsection*{Summary of our framework}
The application of the AGSA method requires the specification of three inputs. The first input is the \textit{contact matrix} (see Section \ref{sec:method_CM}) describing the interaction structure of the population in question.
The second input is the \textit{age-structured epidemic model}, for which we aim to analyze the sensitivity of the contact matrix elements concerning appropriately defined objective functions.
Such objective functions may include the number of infected individuals, those requiring intensive care at the peak of the epidemic wave, or the cumulative number of deaths over the entire epidemic.
The third input consists of the additional \textit{parameters of the model}.

In the first step, the elements of the contact matrix are sampled using the Latin Hypercube Sampling (LHS) method (see Section \ref{sec:method_LHS}). 
For each sampled element of the population, denoted by $\mathcal{S}$, and the fixed additional model parameters, $\mathcal{S}$ simulations are performed. 
From these simulations, the values of the objective functions are determined and paired with the corresponding sample element.

\begin{figure}[h]
    \centering
\includegraphics[trim={5cm 4cm 4cm 6cm},clip,width=\textwidth]
{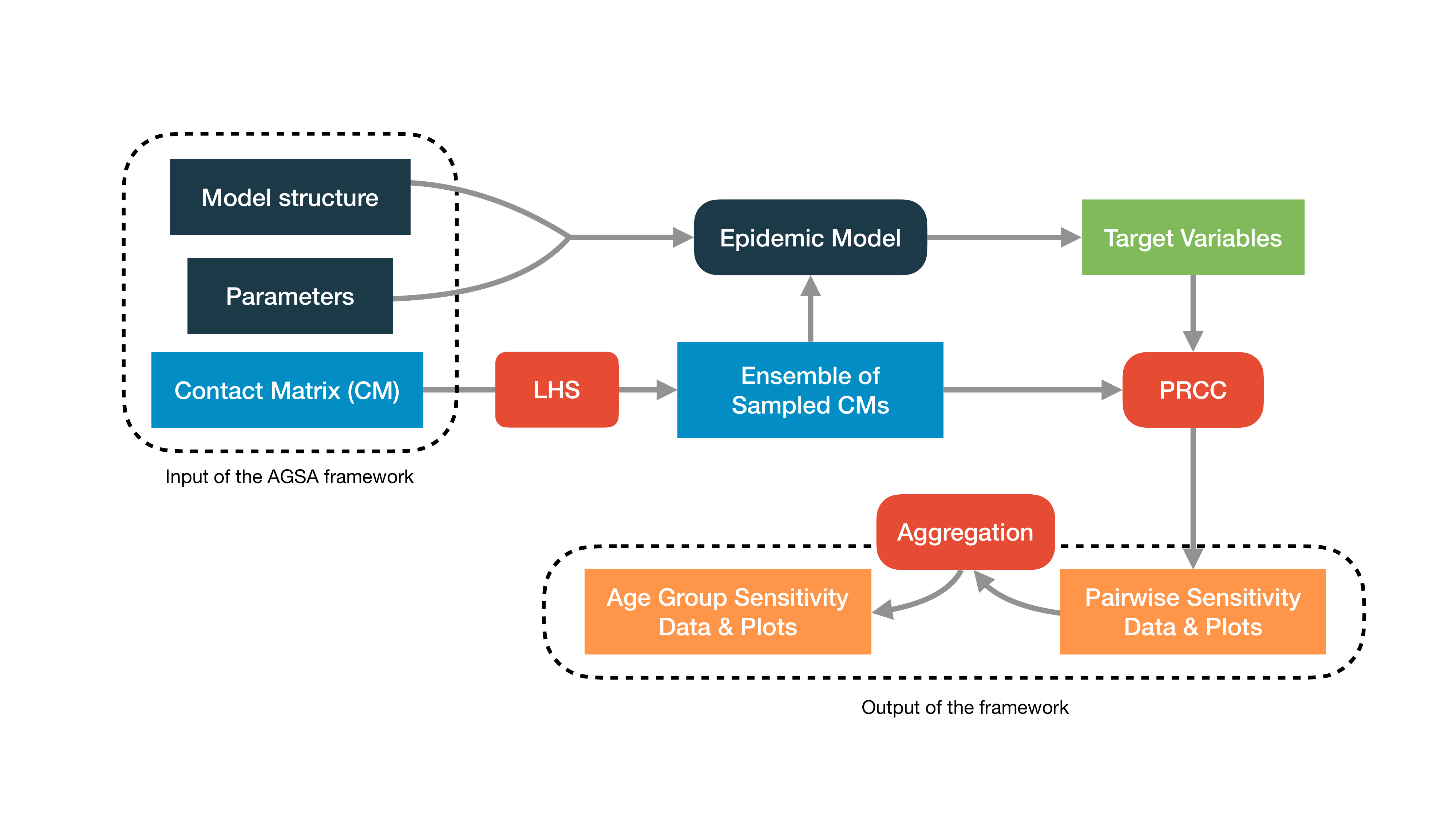}
\caption{\textbf{Workflow of the AGSA framework.} The input parameters include the contact matrix (CM) describing the interaction structure of the population and additional model parameters. Using Latin Hypercube Sampling (LHS), an ensemble of sampled contact matrices is generated and used with the epidemic model to calculate target variable values. Partial Rank Correlation Coefficient (PRCC) analysis is performed on these results to derive sensitivity measures, which are then aggregated by age group to determine group-specific sensitivities. Outputs include sensitivity data and visualizations.}
\label{fig: pipeline}
\end{figure}

Subsequently, PRCC values are calculated (see Section \ref{sec:method_PRCC} for the contact matrix samples and the corresponding objective function values, from which sensitivity measures are derived. 
Finally, the obtained sensitivity measures are aggregated (see Section \ref{sec:method_aggregation}) by age group, yielding the sensitivity of each age group concerning the target variables.
The flowchart of the complete framework is illustrated in Fig. \ref{fig: pipeline}.

The framework was developed from the ground up in Python using an object-oriented programming approach. This ensures ease of extension and modification to accommodate a wide range of applications. 
The implementation utilizes core Python libraries, including \textit{numpy}, \textit{scipy}, and \textit{smt}, to perform the necessary calculations efficiently.

\subsection{Social contact matrices}\label{sec:method_CM}
In Sec. Demonstrations of the present paper we use two social contact matrices: in the case of the SEIR model, the CM is by Mosson et al. \cite{mossong2008social}, but in the case of the Covid model, the CM is derived from Prem et al. \cite{prem2021projecting}.
In the following, we briefly introduce the latter one.
The authors obtained the contact structures of populations in 177 geographical locations. 
The population of one such location is stratified into $n=16$ age groups. 
Except for the last age group, which is dedicated to the elderly population, each age group spans five years: 0-4, 5-9, 10-14, $\dots$, 75+, with $N_i$ representing the number of people in age group $i \in \{0,\dots,15\}$.

\begin{figure}[!ht]
    \centering
    \begin{subfigure}{.19\textwidth}
      \centering
\includegraphics[width=1\linewidth]{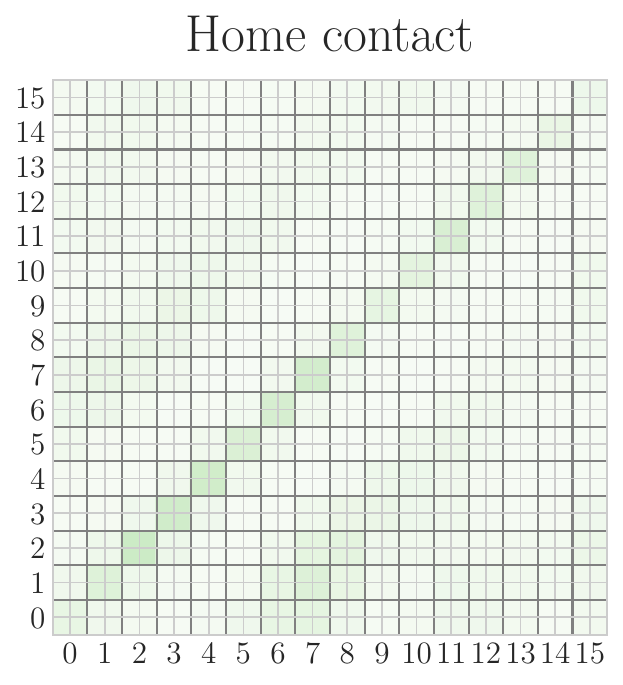}
      \label{fig:sub1}
    \end{subfigure}%
    \begin{subfigure}{.19\textwidth}
      \centering
\includegraphics[width=1\linewidth]{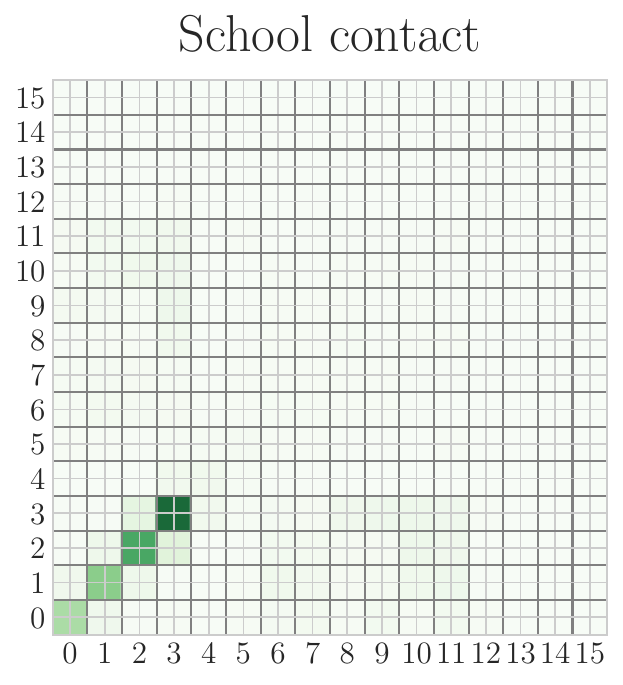}
      \label{fig:sub2}
    \end{subfigure}
    \begin{subfigure}{.19\textwidth}
      \centering
      \includegraphics[width=1\linewidth]{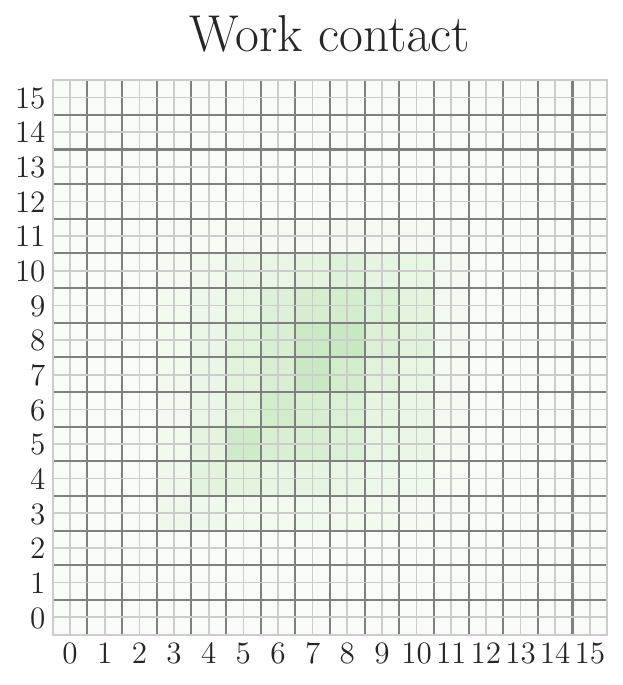}
      \label{fig:sub3}
    \end{subfigure}
    \begin{subfigure}{.19\textwidth}
      \centering
      \includegraphics[width=1\linewidth]{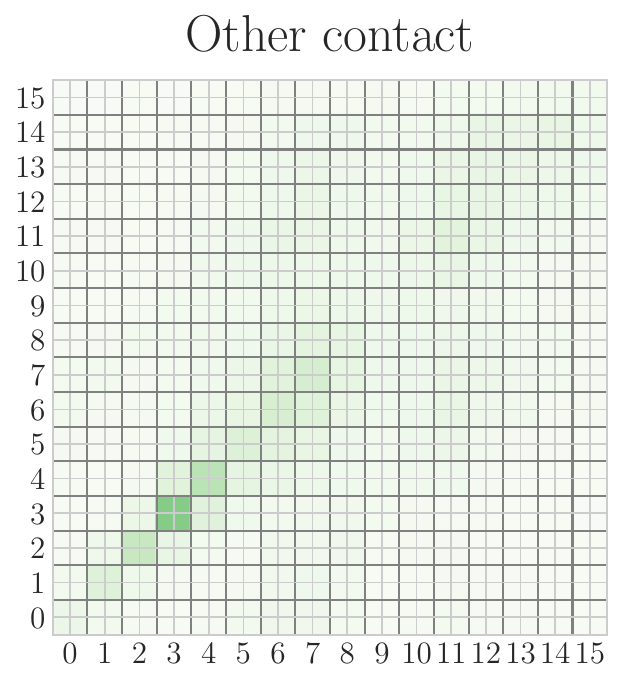}
      \label{fig:sub4}
    \end{subfigure}
    \begin{subfigure}{.19\textwidth}
      \centering
    \includegraphics[width=1.\linewidth]{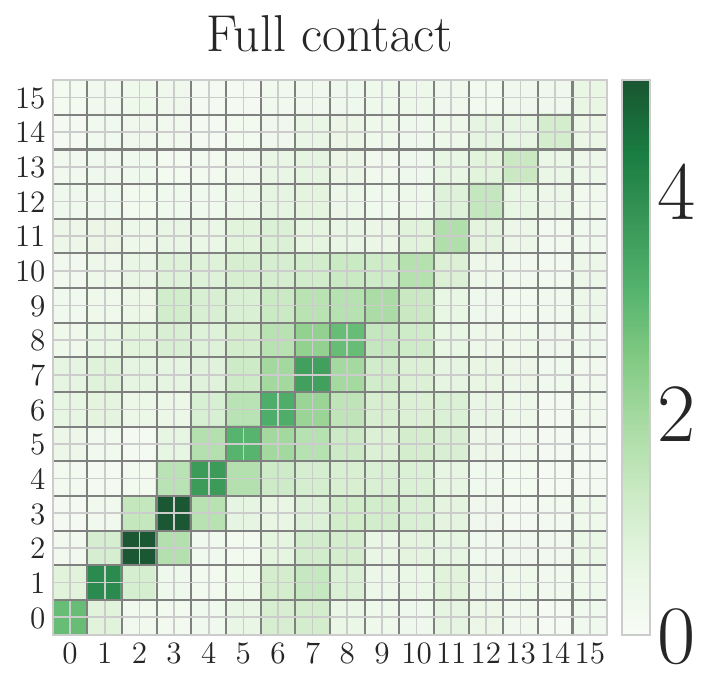}
    \label{fig:sub5}
    \end{subfigure}
    \caption{Contact matrices representing the social contact patterns in Hungary, estimated in \cite{prem2021projecting}. The full contact matrix is the sum of the available, environment-specific matrices showing the distribution of all contacts among the age groups.}
    \label{fig:test}
    \end{figure}

Social contacts are categorized into four environments: home, school, work, and other (which includes typical social venues such as malls, pubs, public transport, etc.). A dedicated contact matrix for each environment is constructed using statistical techniques, specifically Markov chain Monte Carlo simulations and convergence diagnosis.
Resulting matrix \( \tilde{C}^H \) for \textit{home}, \( \tilde{C}^S \) for \textit{school}, \( \tilde{C}^W \) for \textit{work}, and \( \tilde{C}^O \) for \textit{other} settings.  
The full contact matrix is 
$\tilde{C} = \tilde{C}^\text{H} + \tilde{C}^\text{S} + \tilde{C}^\text{W} + \tilde{C}^\text{O}$. 
Contact matrices from \cite{prem2021projecting} for Hungary are displayed in Fig. \ref{fig:test}. 
The elements of the contact matrices will be considered as parameters in the sensitivity analysis.

In particular, the social contact matrix $\tilde{C} \in \mathbb{R}^{n \times n}$ is asymmetric, as the element $(i, j)$ is calculated for the subpopulation $i$, while the element $(j, i)$ is an average projected onto the age group $j$, leading to differences when population sizes vary \cite{gimma2022changes, prem2017projecting, beraud2015french}.

Formally, for a population vector $\bar N=(N_0,\dots,N_{15})$, we expect a proper contact matrix $C$ to satisfy the following condition: 
\begin{equation}
    C_{i, j} N_i = C_ {j,i} N_j.
    \label{Eq:CM_symmetry_cond}
\end{equation}
Thus, we adopt the symmetrization approach from Knipl et al. \cite{knipl2009influenza} similar to that in \cite{ mossong2008social, korir2022clustering, korir2023clusters, klepac2020contacts, mccarthy2020quantifying, rost2020early} to obtain the corrected version of the \textit{home}, \textit{school}, \textit{work} and \textit{other} contacts.
The elements of these new contact matrices $C^\bullet$ are defined as: 
\begin{equation}
C^\bullet_{i,j} = \frac{\tilde{C}^\bullet_{i,j} N_i + \tilde{C}^\bullet_{j,i} N_j}{2 N_i}.
\label{eq3}
\end{equation}
Note that the obtained contact matrices $C^\bullet$ may not be symmetric but fulfill the required symmetry condition (Eq. \ref{Eq:CM_symmetry_cond}).
In our analysis, we use the notation $C=C^H+C^S+C^W+C^O$ and: \begin{equation} 
\overline{C} = [\overline{C}_{i,j}]_{i,j=0}^{15} \in \mathbb{R}^{16 \times 16}, \label{eq4} 
\end{equation} 
where $\overline{C}_{i,j} = C_{i,j} \cdot N_i$,
representing the total number of contacts between age groups. 
This matrix, the \textit{total contact matrix}, is symmetric and contains only $136$ unique elements.

It is important to note that in reality, every contact-related intervention, such as closing schools, maintaining social distance, and working remotely, primarily affects the elements of $C^S, C^W, C^O$. This means that interactions at home or within the household should be treated separately. We use the following notation: $C^{\Delta}$, which represents the sum of $C^\text{S}$, $C^\text{W}$, and $C^\text{O}$, hence, 
\begin{equation}
C^{\Delta} = C^\text{S}+C^\text{W}+C^\text{O}
\label{eq:C_Delta} 
\end{equation}
Furthermore, it is clear that $C=C^{\text{H}} + C^{\Delta}$. The elements of the contact matrices will be considered as parameters in the sensitivity analysis.

\subsection{Age stratified epidemic models and target variables}

The AGSA method can be applied to any age-structured epidemic model to analyze the sensitivity of age groups' impacts on the model's target variables.
These target variables may include any model output, such as the cumulative number of deceased individuals throughout the epidemic, the peak hospitalization rate, or the ICU bed demand.
We demonstrate our method on two age-stratified epidemic models with different complexity.
The most important features of these models are briefly introduced in Section \ref{sec:demo}.

A brief note on using the basic reproduction number as a target variable. To determine the model's base transmission rate, we assume the \( \mathcal{R}_0 \) value the disease would have in the absence of interventions. 
Given this predefined \( \mathcal{R}_0 \), along with the model parameters and the contact matrix, the base transmission rate \( \beta_0 \) can be calculated. 
In real-life situations, contact-based interventions do not change this \( \beta_0 \) as it is a parameter characteristic of the disease. 

During sensitivity analysis, modifying the contact matrix elements alters \( \mathcal{R}_0 \) since the next-generation matrix (see Sec. \ref{sec:method_LHS}) computation incorporates these elements. 
As a result, \( \mathcal{R}_0 \) can also be selected as a target variable, allowing us to assess how changes in contact patterns influence the basic reproduction number.

\subsection{Latin Hypercube Sampling}\label{sec:method_LHS}

Latin Hypercube Sampling (LHS) is a statistical technique designed to generate a representative sample of parameters from a multidimensional distribution. Unlike simple random sampling, LHS achieves comparable or superior results with a smaller sample size, making it a more efficient alternative. The method was introduced by Michael McKay et al. in 1979 \cite{mckay2000comparison}.

LHS assumes that each parameter is associated with a probability distribution and a corresponding density function \cite{blower1994sensitivity, mckay1992latin}. 
According to McKay, the sample size $\mathcal{S}$ should satisfy $\mathcal{S}> \frac{4}{3} K$, where $K$ represents the number of parameters under consideration \cite{mckay2000comparison}.

In the sampling process, the range of each parameter is divided into $\mathcal{S}$ non-overlapping intervals, each with equal probability based on its density function. 
Then, a single interval is selected randomly (without repetition), and a random value is drawn from within that interval. 
This procedure is repeated for all parameters in the analysis.
The resulting LHS matrix has rows that represent individual sample vectors $(p^{(i)}_1,\dots, p^{(i)}_K)$, with each vector used as input for a separate simulation. 
The outputs of these simulations are then generated, and the values $v^{(i)}$ of the investigated target variable are obtained for every parameter configuration. This is illustrated in Table \ref{table:LHS}.

\begin{table}[!ht]
\centering
\begin{tabular}{|c|c|c|c||c|c|c|c|}
    \hline
    \rowcolor[HTML]{C0C0C0}
    \multicolumn{4}{|c||}{LHS Parameter Samples} & Target Variable \\ \hline
    $p_{1}$ & $p_{2}$ & $\cdots$ & $p_{K}$ & $v$ \\ \hline
    $p_{1}^{(1)}$ & $p_{2}^{(1)}$ & $\cdots$ & $p_{K}^{(1)}$ & $v^{(1)}$ \\ 
    $p_{1}^{(2)}$ & $p_{2}^{(2)}$ & $\cdots$ & $p_{K}^{(2)}$ & $v^{(2)}$ \\ 
    $\vdots$ & $\vdots$ & $\ddots$ & $\vdots$ & $\vdots$ \\ 
    $p_{1}^{(\mathcal{S})}$ & $p_{2}^{(\mathcal{S})}$ & $\cdots$ & $p_{K}^{(\mathcal{S})}$ & $v^{(\mathcal{S})}$ \\ \hline
\end{tabular}
\caption{
\textbf{Structure of the LHS matrix and its associated outputs.} 
The left section illustrates the parameter samples \( p_k^{(i)} \), the \( i \)-th sample of the \( k \)-th parameter, generated for each of the \( \mathcal{S} \) sample vectors. The right section presents the corresponding value \( v^{(i)} \) of the target variable, obtained from the simulation performed using the \( i \)-th sample vector. Each row represents a unique sample-output pair.}
\label{table:LHS}    
\end{table}

\subsubsection*{Sampling the Contact Matrix}
Upon symmetrization, we have to sample the lower triangular elements of the CM.
That is 136 independent parameters in the case of the CM introduced in Sec. \ref{sec:method_CM}.
To ensure the sampling process is epidemiologically relevant, we make the following assumptions:
\begin{enumerate}[\bfseries {A}1]
    \item We assume that no intervention can alter home contacts since a complete lockdown restricts individuals to their households.  
    Thus, we should only sample the changes for ${C}^{\Delta}={C}^W+{C}^S+{C}^O$ (see Eq. \ref{eq:C_Delta}).
    \item The target parameters for our PRCC analysis include the reproduction numbers $\mathcal{R}_0$, the infected peak, the hospitalized peak, the peak of the intensive care unit, and the final death size. 
    Therefore, to assess the effect of variations on outbreaks, we must ensure that our sampling keeps $\mathcal{R}_0 \geq 1$. 
    Thus, we only consider sampled contact matrices for which the basic reproduction number exceeds 1.
    \item We aim to examine the overall sensitivity of all contacts characteristic of a given age group rather than limiting the analysis to specific contacts stored in the workplace, school, or other matrices. 
    Thus, it is more practical to sample the \textit{reduction ratios} of the elements of $C^{\Delta}$ rather than the elements of the contact matrix directly.
\end{enumerate}

To assess \textbf{A2}, first, we obtain the basic reproduction number $\mathcal{R}_0$ using the next generation matrix (NGM \cite{diekmann2010construction}).
Then, we introduce parameter $\kappa$ and sample proportions from $[0, \kappa]$. 
Our objective is to find the value of $\kappa$ so that, $\mathcal{R}_0=1$ with the contact matrix
$$ C^{\text{H}} + (1-\kappa) \cdot C^{\Delta}.$$ 
The appropriate value of $\kappa$ can be identified using basic interval logic or more advanced techniques.

To assess \textbf{A3}, we randomly sample the vector $\bar W$ of $136$ unique elements independently within the interval $[0, \kappa]$ using a uniform distribution.
We use a uniform distribution because we assume no prior knowledge about the distribution of individual contact pairs. If such prior information is available, the appropriate distribution can be selected accordingly.
These elements represent proportional changes and are arranged in a square symmetric matrix,
$$M_{\text{ratio}}\in{[0,\kappa]}^{16\times16}, \quad\text{and it is true that }\quad M_\text{ratio}^{(i,j)} = M_\text{ratio}^{(j,i)}.$$
The mapping between the sampled vector $\bar W$ and $M_\text{ratio}$ can be seen in Appendix \ref{c_mtx_map}.
We adjust the matrices by multiplying elementwise the total matrix ${C}^{\Delta}$ by $\mathbbm{1}-M_{\text{ratio}}$, resulting $$C^{\Delta}_{\text{sampled}}=( \mathbbm{1} - M_{\text{ratio}}) \odot {C}^{\Delta}.$$ 
And the modified total matrix ${C}'$ has the form of
\begin{equation}
        {C}' = {C}^{\text{H}} + C^{\Delta}_{\text{sampled}},
        \label{eq:csample}
\end{equation}
where $\mathbbm{1} \in \mathbb{R}^{16\times 16}$ represents the matrix $[1]_{i,j=1}^{16}$, $\odot$ indicates element-wise multiplication.
In the extreme case, where the maximum change is applied, $M_\text{ratio} = \mathbbm{1}$, resulting in ${C}' = {C}^{\text{H}}$. 
Using equation \ref{eq:csample}, reducing the contacts below the home contacts is impossible.

\subsection{Partial Rank Correlation Coefficient (PRCC)}\label{sec:method_PRCC}

The Partial Rank Correlation Coefficient (PRCC) is a widely used tool for sensitivity analysis based on Latin Hypercube Sampling (LHS). Pearson's correlation coefficient, partial correlation coefficient, and standardized regression coefficients assume linear relationships between variables. In contrast, the rank correlation variants test for non-linear but monotonic relationships. This method replaces actual values with rank numbers, thereby mitigating the effect of non-linearity in the relationship. Examples of rank correlation methods include SRCC (Spearman Rank Correlation Coefficient), PRCC, and SRRC (Standardized Rank Regression Coefficient) \cite{blower1994sensitivity, hamby1994review}.

The initial step in calculating the PRCC values involves replacing each LHS value in Table \ref{table:LHS} with its rank value.
A rank value is an integer if the corresponding data point is unique. 
If multiple identical values exist in the sample, they are assigned the average of their respective ranks.
The ranking is executed for every parameter and the simulation output values, as illustrated in Table \ref{table:LHS_ranks}. 
We use $r_i$ ($i=1,2,\dots,K$) to represent the target variable's input parameters with ranks and $d$.
Subsequently, the procedure involves fitting $2K$ regression models in two rounds: using $d$ as a target in the first round and then fitting models with each $r_i$ as the output variable. 
In the second round of linear regression, the rank parameter $r_k$ is estimated using the other rank parameters \cite{marino2008methodology}. In neither round of the linear regression process is the parameter $r_k$ used as an input variable.
    
\begin{table}[!ht]
\centering
\begin{tabular}{|c|c|c|c||c|c|c|c|}
    \hline
    \rowcolor[HTML]{C0C0C0}
    \multicolumn{4}{|c||}{LHS Parameter Ranks} & Output Ranks \\ \hline
    $r_{1}$ & $r_{2}$ & $\cdots$ & $r_{K}$ & $d$ \\ \hline
    $r_{1}^{(1)}$ & $r_{2}^{(1)}$ & $\cdots$ & $r_{K}^{(1)}$ & $d^{(1)}$\\ 
    $r_{1}^{(2)}$ & $r_{2}^{(2)}$ & $\cdots$ & $r_{K}^{(2)}$ & $d^{(2)}$ \\ 
    $\vdots$ & $\vdots$ & $\ddots$ & $\vdots$ & $\vdots$ \\ 
    $r_{1}^{(\mathcal{S})}$ & $r_{2}^{(\mathcal{S})}$ & $\cdots$ & $r_{K}^{(\mathcal{S})}$ & $d^{(\mathcal{S})}$ \\ \hline
\end{tabular}
\caption {Replacing the sampled values in Table \ref{table:LHS} with their respective ranks. Specifically, the $i$th sample of the $k$th variable $p_k^{(i)}$ in Table \ref{table:LHS} is substituted for its rank $r_k^{(i)}$, determined by organizing the values in the column $p_k$ and identifying the position of $p_k^{(i)}$ in the sorted array.
$T$ represents the number of target variables.}
\label{table:LHS_ranks}
\end{table}

The pairwise PRCC value can be found as the Pearson correlation coefficient between the two computed residuals. Specifically, for the $k$-th pairwise PRCC parameter, it is given by 
    $$
    P_k = \rho_{\mathrm{Res}_{1,k}, \mathrm{Res}_{2,k}} = \frac{\mathrm{Cov}(\mathrm{Res}_{1,k}, \mathrm{Res}_{2,k})}{\sqrt{\mathrm{Var}(\mathrm{Res}_{1,k}) \cdot \mathrm{Var}(\mathrm{Res}_{2,k})}} 
    $$
    where $\mathrm{Var}(\mathrm{Res}_{1,k})$ and $\mathrm{Var}(\mathrm{Res}_{2,k})$ denote the variance of $\mathrm{Res}_{1,k}$ and $\mathrm{Res}_{2,k}$, respectively, and $\mathrm{Cov}(\mathrm{Res}_{1,k}, \mathrm{Res}_{2,k})$ is their covariance. Note that $P_k$ ranges from $-1$ to $1$. To simplify the notation, let's consider a single output, then, 
    $$\mathrm{Res}_{1,k} = d^{(i)} - \hat{d}^{(i)}_k,  i = 1, ..., N.$$ $$\mathrm{Res}_{2,k} = r_k^{(i)} - \hat{r}^{(i)}_k,  i = 1, ..., N.$$ where,
     $$
    \hat{d}^{(i)}_{k} = a_k^{(0)} + \sum \limits^K_{j=1, j\neq k} a_k^{(j)} r^{(i)}_{j}
    , i = 1, ...,   N.$$   
and
    $$\hat{r}^{(i)}    _{k} = b_k^{(0)} + \sum \limits^K_{j=1, j\neq k} b_k^{(j)} r^{(i)}_{j},
    i = 1, ...,   N.$$
    
The magnitude of the PRCC value indicates the strength of the relationship between the two variables: a higher absolute coefficient value signifies a stronger relationship between the variables.
In the case of contact structure sensitivity analysis, reducing a contact matrix element leads to a decrease in the target variable values in all scenarios examined in this study.

\subsection{Inference on PRCCs}
Marino et al. \cite{marino2008methodology} describe a technique to determine the significance of tests that measure whether a pairwise PRCC value $P_k$ significantly differs from zero and whether two pairwise PRCC values differ significantly. As per the authors, if the sample size is $N$ and there are $K$ parameters whose effects are considered when calculating $P_k$, each pairwise PRCC ($P_k$) will yield a value $\mathcal{T}_k$ according to the following statistic:
$$\mathcal{T}_k  
  = P_k \sqrt{ \frac{N - 2 - K}{1 - P_k ^ {2}}} \sim t_{N - 2 - K},
  $$
where $\mathcal{T}_k
$
follows a Students-t distribution with $(N - 2 - K)$ degrees of freedom. The p-values $(p_k)$ can be calculated from $\mathcal{T}_k$. Note that $p_k \in [0, 1]$, where the higher values $(p_k > 0.1)$ are statistically insignificant.

The 136 sensitivity measures for \(P_k\) and their corresponding \(p_k\), are arranged in a square symmetric matrix to be used in the aggregation approach.
A lower triangular of a matrix like this can be seen in Fig. \ref{fig:SEIR_peak_size}.
For each \( P_k \) element, we determine the corresponding \((i,j)\) age group pair to which the sensitivity value belongs. This value is then assigned to the \( P_{i,j} \) and \( P_{j,i} \) elements of the \( P \) matrix.
We repeat this process for the \( p_k \) values as well, assigning each to the corresponding \( p_{i,j} \) and \( p_{j,i} \) elements of the \( p \) matrix.
Consequently, we define the matrices \(P\) and \(p\) as follows:
\begin{equation}\label{eq:PRC_mtx}
P = [P_{i,j}]_{i,j=1}^{16} \quad \text{with} \quad P_{i,j} = P_{j,i}, \quad (1 \leq j \leq i \leq 16)
\end{equation}
as the PRCC matrix and
\begin{equation}\label{eq:p-val_mtx}
p = [p_{i,j}]_{i,j=1}^{16} \quad \text{with} \quad p_{i,j} = p_{j,i}, \quad (1 \leq j \leq i \leq 16)
\end{equation}
as the p-value matrix.

\subsection{PRCC aggregation approach}\label{sec:method_aggregation}

We introduce the methodology for aggregating sensitivity values across age group pairs.
For a given \((i,j)\) contact pair, the sensitivity value \( P_{i,j} \) is associated with the corresponding \( p_{i,j} \) value, which represents the uncertainty of the sensitivity measure. Thus, \( 1 - p_{i,j} \) quantifies the reliability of \( P_{i,j} \).
Now, we construct a probability distribution for age group \( i \) using the values of \( 1 - p_{i,j} \). 

Let \( X_i \) be a discrete random variable that takes values from the set \( X_i \in \{P_{i,0}, \dots, P_{i,15} \} \), with probabilities defined as:
\[
\text{Probability}(X_i = P_{i,j}) = \frac{1 - p_{i, j}}{\sum_{m=0}^{15} (1 - p_{i, m})}, \quad \text{where } j \in \{0,\dots,15\}.
\]
This formulation ensures that sensitivity values with higher reliability (\( 1 - p_{i,j} \)) contribute more to the overall distribution of \( X_i \).
Finally, the aggregated sensitivity value for the \( i \)-th age group, denoted as \( \mathbf{P}_i \), is defined as the median of the distribution \( X_i \):
\[
\mathbf{P}_i = \text{median}(X_i).
\]
This ensures that the aggregated sensitivity measure reflects the central tendency of the most reliable sensitivity values associated with age group \( i \).
Furthermore, let the confidence interval \( \mathcal{CI}_i \) corresponding to the aggregated sensitivity values be defined as:
\[
\mathcal{CI}_i = [Q_1(X_i), Q_3(X_i)],
\]
where \( Q_1(X_i) \) is the lower quartile and \( Q_3(X_i) \) is the upper quartile of \( X_i \). 

\section{Demonstrations}\label{sec:demo}
We applied the AGSA method to two age-structured epidemic models of different complexity to illustrate its utility and versatility. 
This section briefly introduces these models and presents the aggregated sensitivity values for various target parameters.
For further details of the models, we refer the reader to the original papers \cite{rost2020early, pitman2012estimating}.
Using the LHS method, the parameter intervals were divided into 10,000 parts, representing the sample size.
Simulations were conducted until the number of infected individuals dropped below one. 
During the simulations, we considered two scenarios: a mild epidemic with ${\mathcal{R}}_0=1.2$ and a severe outbreak with ${\mathcal{R}}_0=2.5$.

\subsection{SEIR model for an influenza epidemic by Pitman et al.}

The first model is an SEIRS (Susceptible-Exposed-Infectious-Recovered-Susceptible) model designed to simulate the spread of influenza, incorporating age-specific mixing patterns in England and Wales \cite{pitman2012estimating}. 
The model employs a \textit{Who Acquires Infection From Whom} (WAIFW) matrix, derived from the POLYMOD study \cite{mossong2008social}, which categorizes interactions among 15 age groups while distinguishing between physical and non-physical contacts. 
This model emphasizes the critical role of pediatric vaccination in mitigating the influenza burden and demonstrates how herd immunity can effectively protect vulnerable populations.
In our simulations, we used the parameters of the original model (along with the population data of England and Wales), with the following modification: we assume no waining of immunity, thus our variant is an SEIR type model and we assume no vaccination in these simulations.

\subsection*{Output of the AGSA method: pairwise PRCC and p values and the age group sensitivity}
\begin{figure}[h!]
    \centering
    \includegraphics[trim={0cm 0cm 26.5cm 0cm},clip,width=0.9\textwidth]{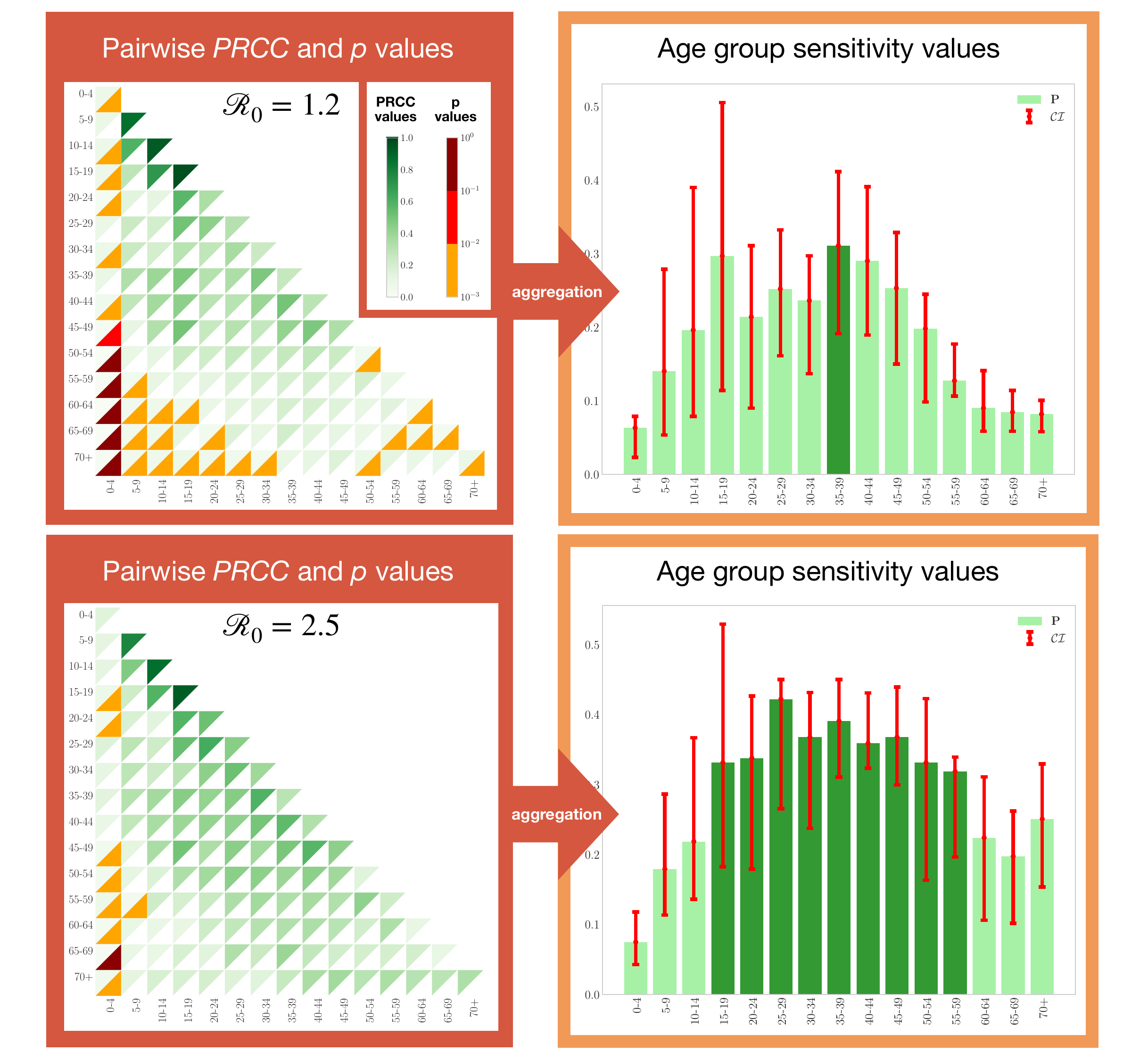}
    \caption{Demonstrations of the AGSA framework in the Influenza model by Pitman et al. \textbf{Target parameter:} number of infected individuals at the epidemic's peak.
    \textbf{Top row:} mild influenza outbreak with $\mathcal{R}_0=1.2$, \textbf{bottom row:} severe outbreak with $\mathcal{R}_0=2.5$.
    In both cases, the target variable is the number of infected individuals at the epidemic's peak.
    \textbf{Left panels:} pairwise PRCC values and associated p-values for the sensitivity analysis of contact matrix elements. Notice that both the diagonal of the PRCC matrix and the p-value matrix (Eq. \ref{eq:PRC_mtx} and \ref{eq:p-val_mtx}) are shown on these panels.
    \textbf{Right panels:} aggregated PRCC values by age group.
    }
    \label{fig:SEIR_peak_size}
\end{figure}

In this section, we demonstrate the output of the contact matrix sensitivity analysis and the output of the aggregation in Fig. \ref{fig:SEIR_peak_size}. 
The left panels of this figure show the pairwise PRCC and $p$ values (according to Eq. \ref{eq:PRC_mtx} and \ref{eq:p-val_mtx}) that are the output of the contact matrix sensitivity analysis. 
Upon aggregation, the right panel highlights age groups with the highest aggregated sensitivity values, indicating their significant influence on the model's target variables. 
In the case of a mild epidemic, the system is most sensitive to contacts of individuals aged 35–39.
However, in the case of a severe epidemic, the contacts of nine age groups (15 to 59) are significant.
This insight demonstrates that, for a given target variable, the sensitivity of different age groups can vary significantly depending on the epidemic's severity.

\subsection{Covid-19 model by Röst et al.}\label{demo:Covid}

The model \cite{rost2020early} we use for the demonstrations is a detailed age-structured COVID-19 model developed for Hungary, incorporating Erlang-like distribution of latent and infectious period, various types of hospitalization, and death compartments. 
The contact data for Hungary is from Prem et al.\cite{prem2017projecting}, and the population vector is obtained from the Hungary Central Statistical Office (KSH).
The model incorporates 16 age groups and 15 specific disease-related classes per age group.

\subsection*{Demonstrations in case of multiple target functions}
\begin{figure}[h!]
    \centering
    \includegraphics[trim={0cm 0cm 0cm 0cm},clip,width=1\textwidth]{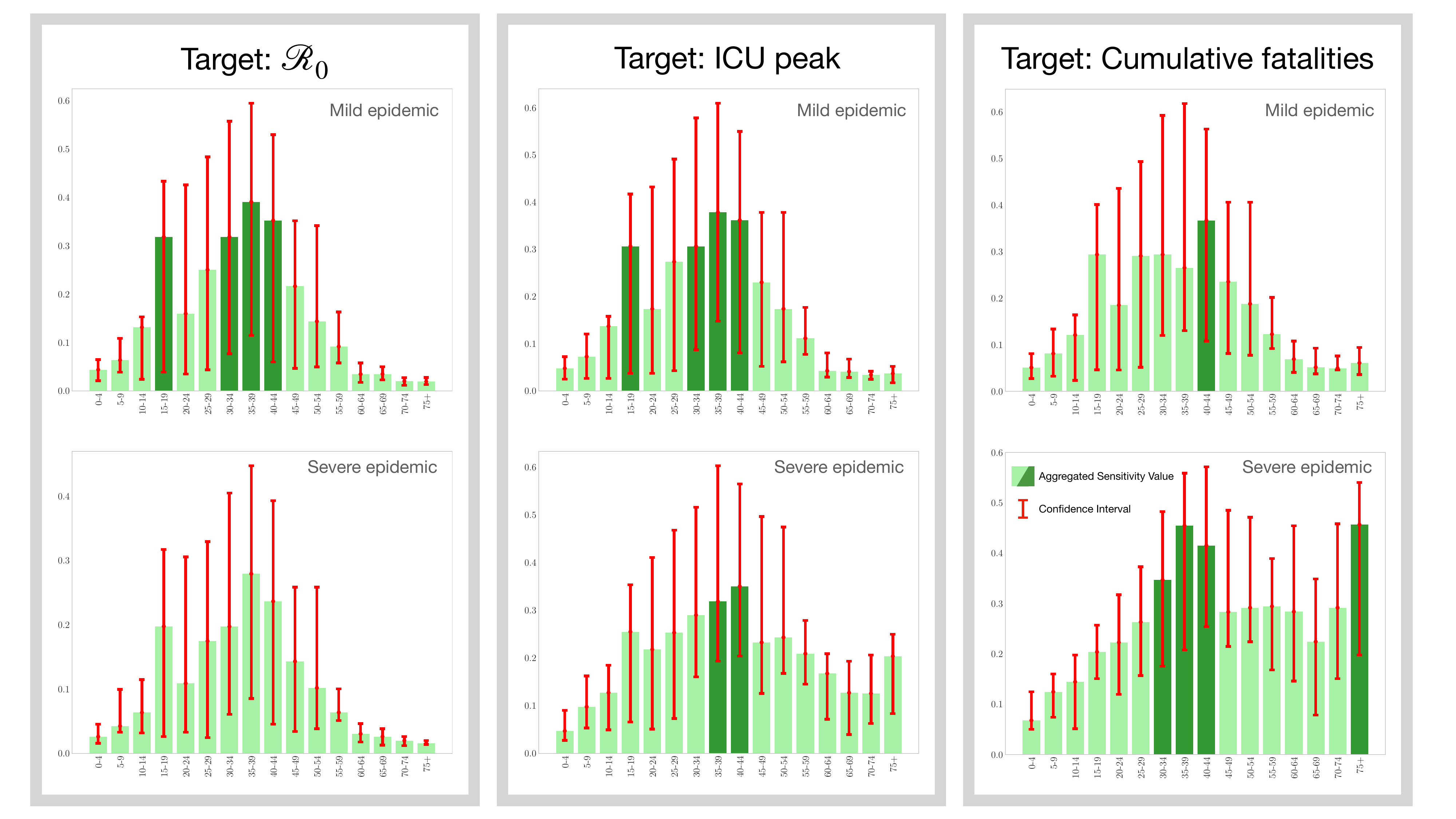}
    \caption{
    Aggregated sensitivity values (\( P \)) and corresponding confidence intervals (CI) across age groups for different target variables in mild (\( \mathcal{R}_0 = 1.2 \), top row) and severe (\( \mathcal{R}_0 = 2.5 \), bottom row) epidemic scenarios. 
    \textbf{The target variables} include the basic reproduction number (\( \mathcal{R}_0 \)), ICU peak occupancy, and cumulative fatalities.
    The legend on the bottom right figure applies to all.}
\label{fig:demo_Rost_many_target}
\end{figure}

This model is sufficiently complex to define several target functions. We analyze the sensitivity of contacts concerning the basic reproduction number, ICU peak, and cumulative fatalities under both mild and severe epidemic scenarios. 
The summarized results are presented in Fig. \ref{fig:demo_Rost_many_target}.

In the case of the \textit{mild epidemic scenario} (\( \mathcal{R}_0 = 1.2 \)), the highest sensitivity values are associated with the 30–44 age groups where the target is the reproduction number and the ICU peak. 
A significant sensitivity is also observed in the 15–19 age group. 
This highlights the critical role of high school-aged individuals in driving transmission dynamics in mild outbreaks, as they have high contact rates in school and social settings.
Interestingly, the number of cumulative fatalities during the whole epidemic is most sensitive for the age group 40-44, not for the elderly age group (75+).

In the \textit{severe epidemic scenario} (\( \mathcal{R}_0 = 2.5 \)):
The 30–44 age group remains the most influential across all target variables, showing consistently high sensitivity values. This age group likely plays a pivotal role due to its high contact rates across various settings, such as workplaces and social environments.
Unlike the mild scenario, the sensitivity of the 15–19 age group significantly diminishes, suggesting that reducing contacts within younger populations has a much smaller impact on epidemic outcomes during a severe epidemic.
Not surprisingly, cumulative fatalities are strongly influenced by contacts in the 75+ age group in case of a severe epidemic scenario.

The contrast between mild and severe epidemic scenarios reveals that
in a \textit{mild outbreak}, reducing contacts in both younger age groups (mainly 15–19) and middle-aged adults (30–44) can effectively mitigate transmission and reduce key outcome measures, including ICU occupancy and cumulative fatalities.
In a severe outbreak, the system becomes less responsive to contact reductions in younger age groups. 
The effectiveness of interventions is primarily driven by targeting middle-aged adults (30–44 years), reflecting their significant role in maintaining transmission chains in high-transmissibility scenarios.

These findings suggest that age-specific intervention strategies should be adapted based on the outbreak's severity.
The figures underscore the importance of tailoring intervention strategies to the disease's transmissibility and highlight the potential of targeted contact reductions to effectively minimize hospital burden and fatalities.

\section{Discussion}

In this study, we presented the Age Group Sensitivity Analysis method, a novel framework designed to quantify the sensitivity of epidemic outcomes to variations in social contact structures across age groups.
The proposed sensitivity aggregation technique provides a nuanced view by highlighting age groups that contribute the most to uncertainty in model predictions, which can be particularly valuable for targeted intervention strategies.
Our findings in Sec. Demonstrations emphasize the importance of tailoring intervention strategies to the epidemic's severity. 

A critical implication of this research lies in its potential application to improve data collection efforts. 
Given the challenges associated with accurately estimating social contact matrices—mainly due to varying survey responses and accessibility across age groups—the AGSA method provides a data-driven basis for focusing empirical studies on age groups that introduce the greatest epistemic uncertainty. 
Concentrating on these influential groups makes it possible to refine contact structure estimates, thus enhancing model predictions.

Furthermore, the framework's ability to incorporate various target variables — peak hospitalization, ICU occupancy, and cumulative fatalities — demonstrates its versatility and practical utility for public health decision-making. Policymakers can leverage the insights provided by AGSA to prioritize interventions that align with specific epidemic control objectives, whether the focus is reducing transmission rates, minimizing hospital burden, or lowering mortality.

The AGSA framework offers a comprehensive and adaptable approach to understand the role of age-specific contact structures in disease transmission dynamics. 
This method can inform more precise and effective public health responses by identifying the most influential age groups across different epidemic scenarios. 
Future research could extend this framework by incorporating other demographic factors, such as socioeconomic status or health comorbidities, further enhancing its applicability to real-world epidemic management.

\section{Ethics Statement}
\textbf{Declaration of generative AI and AI-assisted technologies in the writing process
}\\During the preparation of this work, the authors used ChatGPT-4o to enhance clarity, grammar, and readability.
After using this tool, the authors reviewed and edited the content as needed and take full responsibility for the content of the publication.

\section{Data Availability}
Code and sensitivity-related data used for this analysis are available on Github: \url{https://github.com/zsvizi/sensitivity-contact-epidemics}

\section{Acknowledgements}
This project was partially supported by the National Laboratory for Health Security Program (RRF-2.3.1-21-2022-00006) and the NKFIH KKP 129877, both funded by the Ministry of Innovation and Technology of Hungary through the National Research, Development, and Innovation Fund.

This research was partially supported by the project TKP2021-NVA-09, provided by the Ministry of Culture and Innovation of Hungary from the National Research, Development and Innovation Fund, financed under the TKP2021-NVA funding scheme.

 \bibliographystyle{elsarticle-num} 
 \bibliography{AGSA}

\begin{thebibliography}{10}
\expandafter\ifx\csname url\endcsname\relax
  \def\url#1{\texttt{#1}}\fi
\expandafter\ifx\csname urlprefix\endcsname\relax\def\urlprefix{URL }\fi
\expandafter\ifx\csname href\endcsname\relax
  \def\href#1#2{#2} \def\path#1{#1}\fi

\bibitem{alam2020parameter}
M.~Alam, M.~Kamrujjaman, M.~Islam, et~al., Parameter sensitivity and
  qualitative analysis of dynamics of ovarian tumor growth model with treatment
  strategy, Journal of Applied Mathematics and Physics 8~(06) (2020) 941--955.

\bibitem{wu2013sensitivity}
J.~Wu, R.~Dhingra, M.~Gambhir, J.~V. Remais, Sensitivity analysis of infectious
  disease models: methods, advances and their application, Journal of The Royal
  Society Interface 10~(86) (2013) 20121018.

\bibitem{blower1994sensitivity}
S.~M. Blower, H.~Dowlatabadi, Sensitivity and uncertainty analysis of complex
  models of disease transmission: an hiv model, as an example, International
  Statistical Review/Revue Internationale de Statistique (1994) 229--243.

\bibitem{nsoesie2012sensitivity}
E.~O. Nsoesie, R.~J. Beckman, M.~V. Marathe, Sensitivity analysis of an
  individual-based model for simulation of influenza epidemics, PloS one 7~(10)
  (2012) e45414.

\bibitem{marino2008methodology}
S.~Marino, I.~B. Hogue, C.~J. Ray, D.~E. Kirschner, A methodology for
  performing global uncertainty and sensitivity analysis in systems biology,
  Journal of theoretical biology 254~(1) (2008) 178--196.

\bibitem{hamby1994review}
D.~M. Hamby, A review of techniques for parameter sensitivity analysis of
  environmental models, Environmental monitoring and assessment 32 (1994)
  135--154.

\bibitem{sobol1990sensitivity}
I.~M. Sobol', On sensitivity estimation for nonlinear mathematical models,
  Matematicheskoe modelirovanie 2~(1) (1990) 112--118.

\bibitem{morris1991factorial}
M.~D. Morris, Factorial sampling plans for preliminary computational
  experiments, Technometrics 33~(2) (1991) 161--174.

\bibitem{friendly2005early}
M.~Friendly, D.~Denis, The early origins and development of the scatterplot,
  Journal of the History of the Behavioral Sciences 41~(2) (2005) 103--130.

\bibitem{helton2000sampling}
J.~C. Helton, F.~J. Davis, Sampling-based methods for uncertainty and
  sensitivity analysis, Tech. rep., Sandia National Lab.(SNL-NM), Albuquerque,
  NM (United States); Sandia~… (2000).

\bibitem{gomero2012latin}
B.~Gomero, Latin hypercube sampling and partial rank correlation coefficient
  analysis applied to an optimal control problem, Master's thesis, University
  of Tennessee (2012).

\bibitem{prem2021projecting}
K.~Prem, K.~v. Zandvoort, P.~Klepac, R.~M. Eggo, N.~G. Davies, C.~for the
  Mathematical Modelling~of Infectious Diseases COVID-19 Working~Group, A.~R.
  Cook, M.~Jit, Projecting contact matrices in 177 geographical regions: an
  update and comparison with empirical data for the covid-19 era, PLoS
  computational biology 17~(7) (2021) e1009098.

\bibitem{mossong2008social}
J.~Mossong, N.~Hens, M.~Jit, P.~Beutels, K.~Auranen, R.~Mikolajczyk,
  M.~Massari, S.~Salmaso, G.~S. Tomba, J.~Wallinga, et~al., Social contacts and
  mixing patterns relevant to the spread of infectious diseases, PLoS medicine
  5~(3) (2008) e74.

\bibitem{fumanelli2012inferring}
L.~Fumanelli, M.~Ajelli, P.~Manfredi, A.~Vespignani, S.~Merler, Inferring the
  structure of social contacts from demographic data in the analysis of
  infectious diseases spread, PLOS Computational Biology (2012).

\bibitem{iozzi2010little}
F.~Iozzi, F.~Trusiano, M.~Chinazzi, F.~C. Billari, E.~Zagheni, S.~Merler,
  M.~Ajelli, E.~Del~Fava, P.~Manfredi, Little italy: an agent-based approach to
  the estimation of contact patterns-fitting predicted matrices to serological
  data, PLoS computational biology 6~(12) (2010) e1001021.

\bibitem{le2018characteristics}
O.~Le~Polain~de Waroux, S.~Cohuet, D.~Ndazima, A.~Kucharski, A.~Juan-Giner,
  S.~Flasche, E.~Tumwesigye, R.~Arinaitwe, J.~Mwanga-Amumpaire, Y.~Boum,
  et~al., Characteristics of human encounters and social mixing patterns
  relevant to infectious diseases spread by close contact: a survey in
  southwest uganda, BMC infectious diseases 18~(1) (2018) 1--12.

\bibitem{kiti2014quantifying}
M.~C. Kiti, T.~M. Kinyanjui, D.~C. Koech, P.~K. Munywoki, G.~F. Medley, D.~J.
  Nokes, Quantifying age-related rates of social contact using diaries in a
  rural coastal population of kenya, PloS one 9~(8) (2014) e104786.

\bibitem{ajelli2017estimating}
M.~Ajelli, M.~Litvinova, Estimating contact patterns relevant to the spread of
  infectious diseases in russia, Journal of theoretical biology 419 (2017)
  1--7.

\bibitem{melegaro2017social}
A.~Melegaro, E.~Del~Fava, P.~Poletti, S.~Merler, C.~Nyamukapa, J.~Williams,
  S.~Gregson, P.~Manfredi, Social contact structures and time use patterns in
  the manicaland province of zimbabwe, PloS one 12~(1) (2017) e0170459.

\bibitem{kumar2018interacts}
S.~Kumar, M.~Gosain, H.~Sharma, E.~Swetts, R.~Amarchand, R.~Kumar, K.~E.
  Lafond, F.~S. Dawood, S.~Jain, M.-A. Widdowson, et~al., Who interacts with
  whom? social mixing insights from a rural population in india, PLoS One
  13~(12) (2018) e0209039.

\bibitem{read2014social}
J.~M. Read, J.~Lessler, S.~Riley, S.~Wang, L.~J. Tan, K.~O. Kwok, Y.~Guan,
  C.~Q. Jiang, D.~A. Cummings, Social mixing patterns in rural and urban areas
  of southern china, Proceedings of the Royal Society B: Biological Sciences
  281~(1785) (2014) 20140268.

\bibitem{horby2011social}
P.~Horby, P.~Q. Thai, N.~Hens, N.~T.~T. Yen, L.~Q. Mai, D.~D. Thoang, N.~M.
  Linh, N.~T. Huong, N.~Alexander, W.~J. Edmunds, et~al., Social contact
  patterns in vietnam and implications for the control of infectious diseases,
  PloS one 6~(2) (2011) e16965.

\bibitem{grijalva2015household}
C.~G. Grijalva, N.~Goeyvaerts, H.~Verastegui, K.~M. Edwards, A.~I. Gil, C.~F.
  Lanata, N.~Hens, R.~P. project, A household-based study of contact networks
  relevant for the spread of infectious diseases in the highlands of peru, PloS
  one 10~(3) (2015) e0118457.

\bibitem{Koltai_reconst}
J.~Koltai, O.~Vásárhelyi, G.~Röst, M.~Karsai, Reconstructing social mixing
  patterns via weighted contact matrices from online and representative
  surveys, Scientific Reports 12~(4690) (2022).

\bibitem{korir2022clustering}
E.~K. Korir, Z.~Vizi, Clustering of countries based on the associated social
  contact patterns in epidemiological modelling, arXiv preprint
  arXiv:2211.06426 (2022).

\bibitem{gimma2022changes}
A.~Gimma, J.~D. Munday, K.~L. Wong, P.~Coletti, K.~van Zandvoort, K.~Prem,
  C.~C.-. working group, P.~Klepac, G.~J. Rubin, S.~Funk, et~al., Changes in
  social contacts in england during the covid-19 pandemic between march 2020
  and march 2021 as measured by the comix survey: A repeated cross-sectional
  study, PLoS medicine 19~(3) (2022) e1003907.

\bibitem{prem2017projecting}
K.~Prem, A.~R. Cook, M.~Jit, Projecting social contact matrices in 152
  countries using contact surveys and demographic data, PLoS computational
  biology 13~(9) (2017) e1005697.

\bibitem{beraud2015french}
G.~B{\'e}raud, S.~Kazmercziak, P.~Beutels, D.~Levy-Bruhl, X.~Lenne,
  N.~Mielcarek, Y.~Yazdanpanah, P.-Y. Bo{\"e}lle, N.~Hens, B.~Dervaux, The
  french connection: the first large population-based contact survey in france
  relevant for the spread of infectious diseases, PloS one 10~(7) (2015)
  e0133203.

\bibitem{knipl2009influenza}
D.~H. Knipl, G.~R{\"o}st, Influenza models with wolfram mathematica,
  Interesting Mathematical Problems in Sciences and Everyday Life (2009) 1--24.

\bibitem{korir2023clusters}
E.~K. Korir, Z.~Vizi, Clusters of african countries based on the social
  contacts and associated socioeconomic indicators relevant to the spread of
  the epidemic, arXiv preprint arXiv:2303.17332 (2023).

\bibitem{klepac2020contacts}
P.~Klepac, A.~J. Kucharski, A.~J. Conlan, S.~Kissler, M.~L. Tang, H.~Fry, J.~R.
  Gog, Contacts in context: large-scale setting-specific social mixing matrices
  from the bbc pandemic project, MedRxiv (2020) 2020--02.

\bibitem{mccarthy2020quantifying}
Z.~McCarthy, Y.~Xiao, F.~Scarabel, B.~Tang, N.~L. Bragazzi, K.~Nah, J.~M.
  Heffernan, A.~Asgary, V.~K. Murty, N.~H. Ogden, et~al., Quantifying the shift
  in social contact patterns in response to non-pharmaceutical interventions,
  Journal of Mathematics in Industry 10 (2020) 1--25.

\bibitem{rost2020early}
G.~R{\"o}st, F.~A. Bartha, N.~Bogya, P.~Boldog, A.~D{\'e}nes, T.~Ferenci, K.~J.
  Horv{\'a}th, A.~Juh{\'a}sz, C.~Nagy, T.~Tekeli, et~al., Early phase of the
  covid-19 outbreak in hungary and post-lockdown scenarios, Viruses 12~(7)
  (2020) 708.

\bibitem{mckay2000comparison}
M.~D. McKay, R.~J. Beckman, W.~J. Conover, A comparison of three methods for
  selecting values of input variables in the analysis of output from a computer
  code, Technometrics 42~(1) (2000) 55--61.

\bibitem{mckay1992latin}
M.~D. McKay, Latin hypercube sampling as a tool in uncertainty analysis of
  computer models, in: Proceedings of the 24th conference on Winter simulation,
  1992, pp. 557--564.

\bibitem{diekmann2010construction}
O.~Diekmann, J.~Heesterbeek, M.~G. Roberts, The construction of next-generation
  matrices for compartmental epidemic models, Journal of the royal society
  interface 7~(47) (2010) 873--885.

\bibitem{pitman2012estimating}
R.~Pitman, L.~White, M.~Sculpher, Estimating the clinical impact of introducing
  paediatric influenza vaccination in england and wales, Vaccine 30~(6) (2012)
  1208--1224.

\end{thebibliography}






\appendix

\section{The Governing Equations of the SEIR Epidemic Model by Pitman et al.}
For age group $i\in \{0,\dots14\}$ the governing equations of the SEIR disease model adopted from \cite{pitman2012estimating} take the form of

\begin{equation}
\begin{aligned}
\frac{dS_i}{dt} &= -\beta_0 S_i I_i, \\
\frac{dE_i}{dt} &= \beta_0 S_i I_i - \sigma E_i, \\
\frac{dI_i}{dt} &= \sigma E_i - \gamma I_i, \\
\frac{dR_i}{dt} &= \gamma I_i.
\end{aligned}
\label{eq: seir}
\end{equation}

\begin{table}[!ht]
    \centering
    \captionsetup{justification=centering} 
    \small 
    \begin{tabular}{p{3cm} p{7.5cm}} 
        \toprule
        \textbf{Variable} & \textbf{Description} \\
        \midrule
        \rowcolor[HTML]{EFEFEF}
        \(S_i(t)\) & Number of susceptible individuals in age group \(i\) \\
        \rowcolor[HTML]{FFFFFF}
        \(E_i(t)\) & Number of exposed individuals in age group \(i\) \\
        \rowcolor[HTML]{EFEFEF}
        \(I_i(t)\) & Number of infectious individuals in age group \(i\) \\
        \rowcolor[HTML]{FFFFFF}
        \(R_i(t)\) & Number of recovered individuals in age group \(i\) \\
        \bottomrule\\
    \end{tabular}
    
    \begin{tabular}{p{2.cm} p{2.cm} p{10.5cm}}
        \toprule
        \textbf{Parameter} & \textbf{Value} & \textbf{Description} \\
        \midrule
        \rowcolor[HTML]{EFEFEF}
        $\beta_0$ & $3.99 \times 10^{-8}$ & Transmission rate coefficient (constant across all age groups) \\
        $\sigma$ & 0.5 & Rate of latent individuals becoming infectious (constant across all age groups) \\
        \rowcolor[HTML]{EFEFEF}
        $\gamma$ & 0.5 & Recovery rate (constant across all age groups) \\
        \bottomrule
    \end{tabular}
    \caption{State variables and parameters of the model by Pitman et al. \cite{pitman2012estimating} and utilizing the UK population for simulations of Eq \ref{eq: seir}.}
\end{table}

\section{The Governing Equations of the Covid-19 Model by Röst et al.}
\begin{figure}[h!]
    \centering
    \includegraphics[width=1\textwidth]{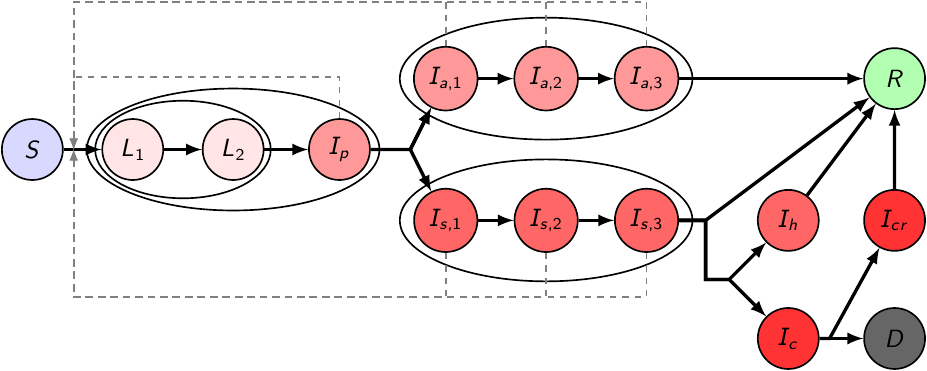}
    \caption{Diagram of the transmission model from \cite{rost2020early} utilized to illustrate the proposed framework. Solid black arrows represent the movement of patients between compartments, whereas dashed arrows show potential pathways for the spread of infection.    
    }
    \label{img:rost_trdiagram}
\end{figure}

In this COVID-19 model (see Fig. \ref{img:rost_trdiagram}), individuals begin in the susceptible compartment (\(S\)), representing those who have not yet been infected but are at risk of exposure. Upon contact with an infectious individual, susceptible individuals transition into the latent phase, progressing through two sequential compartments (\(L_1\) and \(L_2\)). During this period, they are infected but not yet capable of transmitting the disease. After the latent phase, individuals enter the presymptomatic stage (\(I_p\)), where they become infectious but have not yet developed symptoms.

From the presymptomatic phase, individuals diverge into two possible infection pathways. A fraction of them remain asymptomatic, progressing through three stages (\(I_{a,1} \to I_{a,2} \to I_{a,3}\)) before recovering (\(R\)). Despite not showing symptoms, asymptomatic individuals contribute to transmission. The remaining infected individuals develop symptoms and transition through three symptomatic stages (\(I_{s,1} \to I_{s,2} \to I_{s,3}\)). 

A proportion of symptomatic individuals experience severe illness and require hospitalization, moving into the hospitalized compartment (\(I_h\)). From this state, patients may either recover (\(R\)) or deteriorate further, requiring intensive care (\(I_c\)). A fraction of critically ill patients progress to a critical state (\(I_{cr}\)), from which they may either recover (\(R\)) or die from the disease (\(D\)). 

Notations here are aligned with the parameter file located in the repository of the framework. 
Here $\mathrm{inf}_a$ denotes the relative infectiousness of $I_a$ compared to $I_s$, for more details about the other parameters and the methodology for parametrization, see \cite{rost2020early} and the GitHub repository of this project {\small\url{https://github.com/zsvizi/sensitivity-contact-epidemics/blob/master/data/rost_model_parameters.json}}.
The governing equations for age group $i \in \{0, \dots,15\}$ of the Covid model \cite{rost2020early} introduced in Sec. \ref{demo:Covid} take the following form.

 \begin{align}
     {S^i}'(t)={} & -\beta_0 \frac{S^i(t)}{N_i}\cdot \sigma_i \sum_{k=1}^{16} c_{k,i}\left[I_{p}^{k}(t) + \mathrm{inf}_a \sum_{j=1}^3 I_{a,j}^{k}(t) + \sum_{j=1}^3 I_{s,j}^{k}(t)\right] \nonumber\\ 
     {L_1^i}'(t)={} & \beta_0 \frac{S^i(t)}{N_i}\cdot \sigma_i\sum_{k=1}^{16} c_{k,i}\left[I_{p}^{k}(t) + \mathrm{inf}_a \sum_{j=1}^3 I_{a,j}^{k}(t) + \sum_{j=1}^3 I_{s,j}^{k}(t)\right] - 2 \alpha_l L^i_1(t) \nonumber\\
     {L_2^i}'(t)={} & 2 \alpha_l L_1^i(t) - 2\alpha_l L_2^i(t),\nonumber\\
     {I_a^i}'(t)={} & 2 \alpha_l L_{2}^{i}(t) - \alpha_{p} I_{p}^{i} (t)\nonumber\\
     {I_{a,1}^i}'(t)={} & p^{i} \alpha_{p} {I}_{p}^{i} (t) - 3 \gamma_{a} I_{a,1}^{i}(t)\nonumber\\
     {I_{a,2}^i}'(t)={} & 3\gamma_{a} I_{a,1}^{i}(t)- 3\gamma_{a} I_{a,2}^{i}(t)\nonumber\\ 
     {I_{a,3}^i}'(t)={} & 3\gamma_{a} I_{a,2}^{i}(t)- 3\gamma_{a} I_{a,3}^{i}(t) \label{eq:model} \\ 
     {I_{s,1}^i}'(t)= {}& (1 - p^i) \alpha_{p} I_{p}^{i} - 3 \gamma_{s} I_{s,1}^i(t)\nonumber\\
     {I_{s,2}^i}'(t)= {} & 3 \gamma_{s} I_{s,1}^i(t) - 3 \gamma_{s} I_{s,2}^i(t)\nonumber\\
     {I_{s,3}^i}'(t)= {} & 3 \gamma_{s} I_{s,2}^i(t) - 3 \gamma_{s} I_{s,3}^i(t)\nonumber\\
     {I_h^i}'(t)={} & h^i (1 - \xi^i) 3 \gamma_{s} I_{s,3}^i(t) - \gamma_h I_h^i(t)\nonumber\\
     {I_c^i}'(t)={} & h^i \xi^i 3 \gamma_{s} I_{s,3}^i(t)-\gamma_c I_c^i(t)\nonumber\\
     {I_{\mathrm{cr}}^i}'(t)={} &  (1 - \mu^{i}) \gamma_{c} I_{c}^{i} (t) -\gamma_{\mathrm{cr}} I_{\mathrm{cr}}^{i} (t)\nonumber\\
     {R^i}'(t)={} & 3 \gamma_{a} I_{a,3}^{i} (t) + (1- h^{i}) 3 \gamma_{s} I_{s,3}^{i} (t) + \gamma_{h} I_{h}^{i} (t) + \gamma_{\mathrm{cr}} I_{\mathrm{cr}}^{i} (t)\nonumber\\   
     {D^i}'(t)={} & \mu^i \gamma_c I_c^i(t).\nonumber     
    \end{align}

\section{Rearranging the contact matrix elements}\label{c_mtx_map}
We establish a link between the pair \((i, j)\) and \(k\), and then recover \((i, j)\) from \(k\) using specific association rules. These rules are defined as follows:

\begin{itemize}
    \item If \(i = 1\), then \((i, j) \mapsto j\).
    \item If \(i \geq 2\), then \((i, j) \mapsto j + \frac{32 + (2 - i)}{2} \cdot (i - 1)\).
\end{itemize}
To map $k$ to \((i, j)\):

\begin{itemize}
    \item If \(k \leq 16\), then \(k \mapsto (1, j)\).
    \item If \(k > 16\), identify \(i\) such that

    \begin{equation}
    \frac{32 + (2 - i)}{2} \cdot (i - 1) < k < \frac{32 + (2 - (i + 1))}{2} \cdot i.
    \label{i}
    \end{equation}
\end{itemize}
Next, determine \(j\) as
\begin{equation}
j = k - \frac{32 + (2 - i)}{2} \cdot (i - 1)
\label{j}
\end{equation}
for \(1 \leq k \leq 136\).
This method preserves the symmetry of the total contact matrix derived from the adjusted matrix. 

\end{document}